# Universal behavior of giant electroresistance in epitaxial $La_{0.67}Ca_{0.33}MnO_3$ thin films

Y. G. Zhao,[a] Y. H. Wang, G. M. Zhang, B. Zhang, X. P. Zhang, C. X. Yang, P. L. Lang and M. H. Zhu

*Department of Physics, Tsinghua University, Beijing 100084, China*

P. C. Guan

*Department of Physics, National TsingHua University, Hsinchu 300, Taiwan, China*

We report a giant resistance drop induced by dc electrical currents in $La_{0.67}Ca_{0.33}MnO_3$ epitaxial thin films. Resistance of the patterned thin films decreases exponentially with increasing current and a maximum drop shows at the temperature of resistance peak $T_p$. Variation of resistance with current densities can be scaled below and above $T_p$, respectively. This work can be useful for the future applications of electroresistance.

[a] e-mail: ygzhao@tsinghua.edu.cn

The hole doped rare-earth manganites $La_{1-x}A_xMnO_3$, where A is divalent alkaline-earth ions, have become the subject of intense research because of their importance for both fundamental issues in condensed-matter physics and potential applications.[1] It has been shown that spin, charge, and lattice degrees of freedom are strongly coupled in these compounds leading to various properties in the phase diagram of $La_{1-x}A_xMnO_3$.[2-4] For example, $La_{1-x}Ca_xMnO_3$ exhibit ferromagnetic ordering between $0.2<x<0.5$ with a maximum Curie temperature for $x=0.33$. $La_{0.67}Ca_{0.33}MnO_3$ also shows a metal-insulator transition with a peak in the resistance and the effect of magnetoresistace near the Curie temperature $T_c$.[4]

Physical properties of manganites can be tunned by external disturbance due to the breaking balance of various interactions in the material. It has been shown that electric current and/or electric field has strong influence on manganites.[5-14] Most previous work are on the manganites with charge ordered state. Very recently, Gao et al. found that electric current can suppress resistance and shift the temperature of resistance peak $T_p$ of $La_{0.7}Ca_{0.3}MnO_3$ and $La_{0.8}Ca_{0.2}MnO_3$ thin films dramatically.[13, 14] These results are interesting, however, the heating effects of electric current are very strong and make it difficult to observe the intrinsic current effect. For instance, $T_p$ is strongly shifted to low temperatures by the heating effect. Even for the peak resistance, the resistance value can not be determined precisely due to the heating effect, especially in the warming up measurement. In the present letter, we carefully design our experiment in order to minimize the current heating effect. We used the probe to put into liquid nitrogen instead of the cryostat used in ref. [13,14]. Some grease was put between sample and the sample holder to facilitate heat diffusion. Thin films were also patterned with a special design. By doing these, heating effect is almost eliminated. By measuring the temperature dependence of resistance of $La_{0.67}Ca_{0.33}MnO_3$, we found that the thin films show giant electroresistance (ER), and the peak resistance of the samples drops remarkably and displays an exponential decay with current density. However, $T_p$ is not shifted with increasing the electric current, and the intrinsic ER effect can be observed and explained in terms of electric current assisted carrier transport.

$La_{0.67}Ca_{0.33}MnO_3$ (LCMO) thin films were grown on (100) $SrTiO_3$ by pulsed laser deposition using a KrF excimer laser ($\lambda$=248nm). The energy density of the pulsed laser was 1.9J/cm$^2$ with a repetition rate of 8Hz. The oxygen partial pressure was 40Pa, and the substrate temperature was 750°C during deposition. After deposition, LCMO films were cooled down to room temperature in an $O_2$ atmosphere. The thickness of the films is about 100 nm. The phase analysis of the samples was performed using a Rigaku D/max-RB x-ray diffractometer with Cu $k_\alpha$ radiation and the result shows good epitaxy. For the electrical measurement, the films were patterned to get a 50μm wide bridge. Then gold was deposited on the patterned LCMO thin films by magnetron sputtering for contact. The scheme of the patterned film is shown in the inset of Fig.1. The electrical resistance of the samples was measured by the four-probe method and liquid nitrogen was used for the low temperature measurement.

Figure 1 shows the temperature dependence of resistance of LCMO under different dc currents. The warming up and cooling down measurements are consistent as shown in the upper inset of Fig.1, indicating that the heating effect is minor. If the heating effect is severe because of long LCMO bridge and poor heat diffusion, $T_p$ shifts to lower temperatures as shown in Fig.2. This is consistent with the previous result in ref. [13], in which heating problem hinders to see the intrinsic current effect. In this case, the warming up and cooling down measurements show dramatic difference. Fig.1 clearly shows that $T_p$ remains unchanged with increasing the electric current and the peak resistance drops remarkably. The insensitivity of $T_p$ to current differs from the dependence of $T_p$ on magnetic field as the magnetic field shifts $T_p$ to higher temperatures. The value of ER is comparable to magnetoresistance induced by a strong magnetic field (several Tesla).[15] Variation of the peak resistance with dc current density is shown in the inset (b) of Fig.2. By fitting the experimental data, the variation of peak resistance with dc current can be described by $R=2207\exp(-J/1.75)+1899(\Omega)$. Such an exponential decreasing with current is in contrast to the previous studies, where a linear resistance dependence on current is obtained.[13] It should be also pointed out that the value of the resistance comes back up again when the current is reduced, indicating the effect is reversible.

The inset (a) of Fig.2 describes the temperature dependence of ER under different currents. The ER, defined as [R(I)-R(0.001mA)]/ R(0.001mA), is negative because the current reduces the resistance. It is also noted that ER is maximum at $T_p$. ER becomes very small at low temperatures. Figure 3 is the resistance decrease of the sample with current measured at different temperatures. It shows a nonlinear dependence on the current and can be fitted exponentially. For temperatures both above and below $T_p$, the current dependence of resistance shows upward curvature. This behavior is different from the magnetic field dependence of resistance for LCMO,[15] which shows upward curvature bellow $T_p$ and downward curvature above $T_p$.

Markovich et al. studied both ER and MR of a single crystal $La_{0.82}Ca_{0.18}MnO_3$ at four temperatures and found the correlation between ER and MR.[7] A 0.3 mA current has equivalent effect on resistance as 1.5-2 T magnetic field below $T_c$ and 0.4 T at room temperature. Based on the phase separation picture that metallic ferromagnetic clusters embedded in paramagnetic insulating matrix, they explained ER by assuming filamentary currents driven by local electric field gradients which produce very strong local magnetic fields of the order of a Tesla. This scenario may not account for our case, because it will lead to the increase of the transition temperature with current, in conflict with the insensitivity of $T_p$ observed here. Explanations of current induced magnetic field or spin alignment will result in shifting $T_p$ to higher temperatures. It should be pointed out that Wu et al.[11] also found that $T_p$ doesn't change with electric field in their field effect configuration with $La_{0.7}Ca_{0.3}MnO_3$ channel and $PbZr_{0.2}Ti_{0.8}O_3$ gates though their measurement is different form the present work.

Some new idea is needed to explain the ER in the present work. Actually, it is difficult to distinguish the current effect from electric field effect. For ER effect, electric field may play an important role. Hundley et al. found the transport-magnetism correlation in $La_{0.7}Ca_{0.3}MnO_3$ thin film as $\rho(H,T) \propto \exp(-M(H,T)/M_0)$, where $\rho(H,T)$, $M(H,T)$ and $M_0$ are resistivity, magnetization and a constant, respectively.[15] They argued that in manganites, polaron hopping conductivity can be expressed as $\sigma=\sigma_0\exp(-E_a/k_BT)$ in the adiabatic limit with $E_a = W_p - W$, where $W_p$ is a characteristic energy of polarons and W is the bandwidth. In the double exchange model, the charge carrier transfer integral t is an important parameter and is proportional to both W and M. In this way, $\rho(H, T) \propto \exp(-M(H,T)/M_0)$ can be deduced. In our experiment, the effective $E_a$ may decrease due to the extra energy gained by carriers from the electric field. In Fig.4, detailed analysis shows that the current density dependence of R(T,J) can be scaled as $[R(T,J)/R(T,J=0)-1]/A+1=\exp(-kJ/|T-T_p|^{\alpha})$ with $\alpha=0.15$ for $T> T_p$ and $\alpha=0.35$ for $T< T_p$. In analogy to the transport-magnetism correlation $\rho(H,T) \propto \exp(-M(H,T)/M_0)$, a transport-polarization correlation in manganites can be expected $\rho(J,T) \propto \exp(-P(J,T)/P_0)$, where J, P(J,T) and $P_0$ are electric current density, electric polarization and a constant, respectively. From the above transport-polarization correlation, we can further deduce that the electric polarization behaves as $P(J,T) \propto J/|T-T_p|^{\alpha}$. with $\alpha=0.15$ for $T> T_p$ and $\alpha=0.35$ for $T< T_p$. Mayr et al. has studied the resistivity of manganites using a random resistor-network, based on phase separation between metallic and insulating domains.[16] They found when percolation occurs, both as chemical composition or temperature vary, results in good agreement with experiments are obtained. Above $T_p$, resistance is dominated by the insulating domains, while below $T_p$, the effective resistance is determined by the parallel connection of percolational metallic and insulating resistance. This picture may be used to explain the current effect. It is also tempting to use the current induced polarization of non-FM regions to account for the observed effect below $T_p$. In this scenario, the current passing through the FM regions is polarized and then injects into the non-FM insulating regions and keep their polarization within a certain depth, leading to the increase of the FM conducting regions. Therefore the resistance drops. Further work is needed to clarify the underlying mechanism of ER, which will enhance our understanding on manganites.

In summary, a giant ER effect was observed in epitaxial LCMO thin films. The temperature dependence of ER is very similar to that of the resistance versus temperature curve. $T_p$ is not sensitive to electric current. It is also found that the resistance of the samples decreases with current exponentially and scales to two universal lines below and above $T_p$. This work has provided new insights into the manganites, as well as their future applications.

We are grateful to L. Lu for enlightening discussion. This work was supported by NSFC (No. 50272031), the Excellent Young Teacher Program of MOE, 973 project (No. 2002CB613505), and Specialized Research Fund for the Doctoral Program of Higher Education (No. 2003 0003088).

## Figure captions

Fig.1 Temperature dependence of resistance for LCMO under different dc currents. From top to bottom, I=0.001mA, 0.01mA, 0.05mA, 0.1mA, 0.2mA, 0.5mA, 1mA, 2mA, 3mA, 4mA, 5mA and 6mA, respectively. The upper inset shows that the warming up and cooling down measurements are consistent. The lower inset is the scheme of the patterned film. The four grey rectangle parts are deposited gold for contact and the black parts are the patterned LCMO films. The size of the sample is 6mm×3mm and the size of the LCMO bridge is 50μm×50μm.

Fig.2 Temperature dependence of resistance for LCMO under different dc currents for sample with severe heating effect. Inset (a) is the temperature dependence of electroresistance [R(I)-R(0.001mA)]/R(0.001mA) for LCMO under different DC currents without the influence of heating effect (from Fig.1). Inset (b) is the variation of the peak resistance with DC current density (from Fig.1).

Fig.3 Resistance variation with electric current at different temperatures.

Fig.4 Scaling behavior of the current density dependence of R(T, J). The inset shows the temperature dependence of the coefficient A.

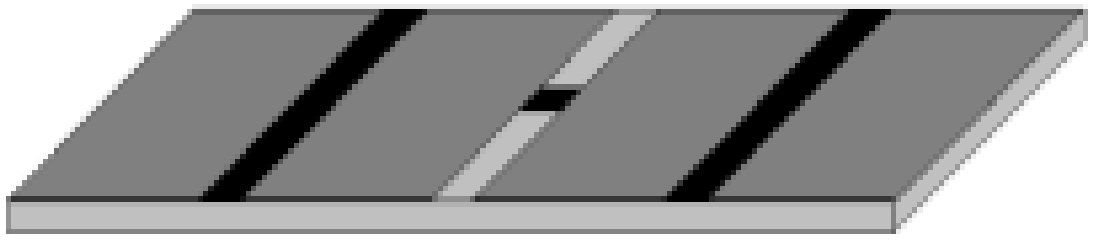

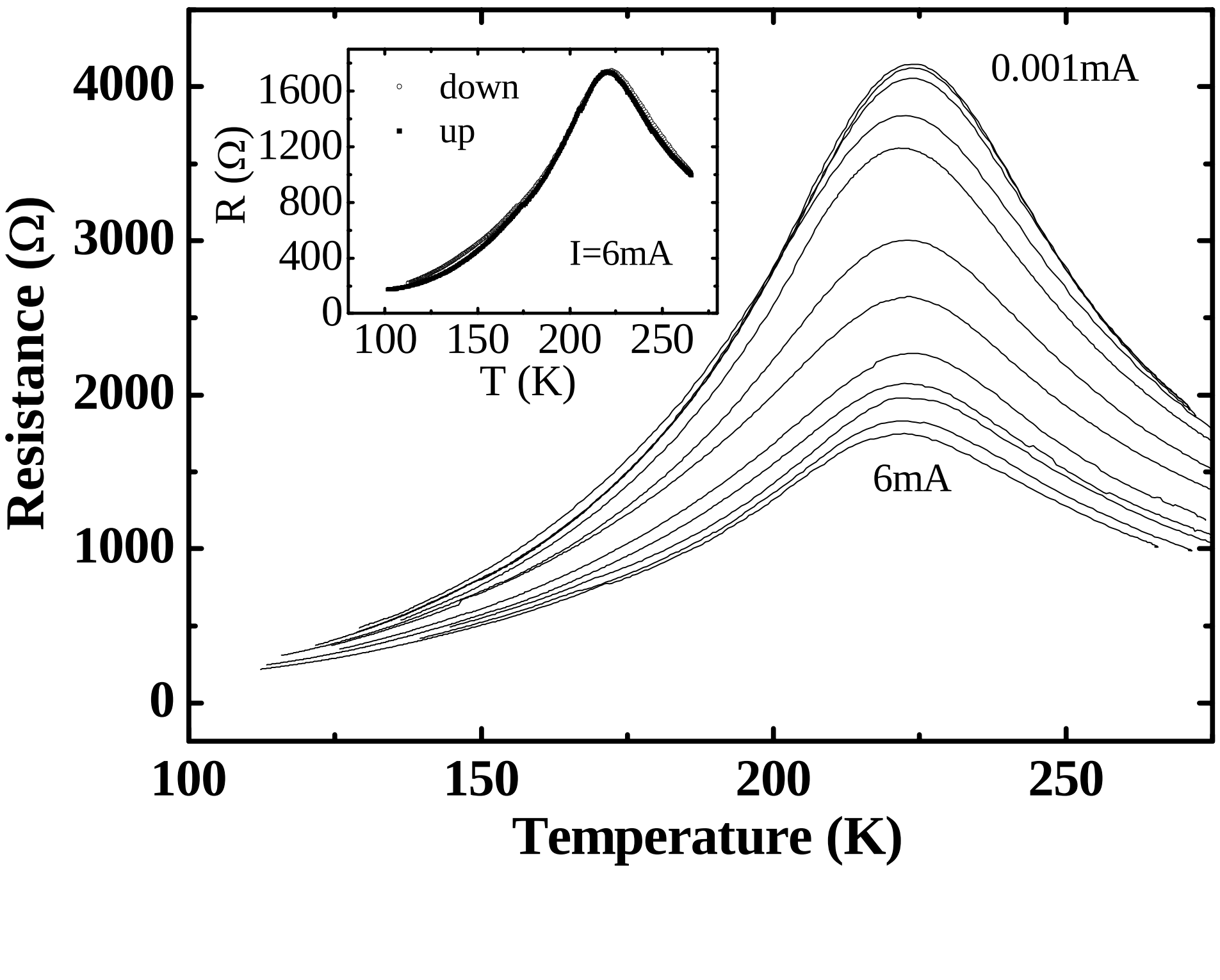

Fig.1

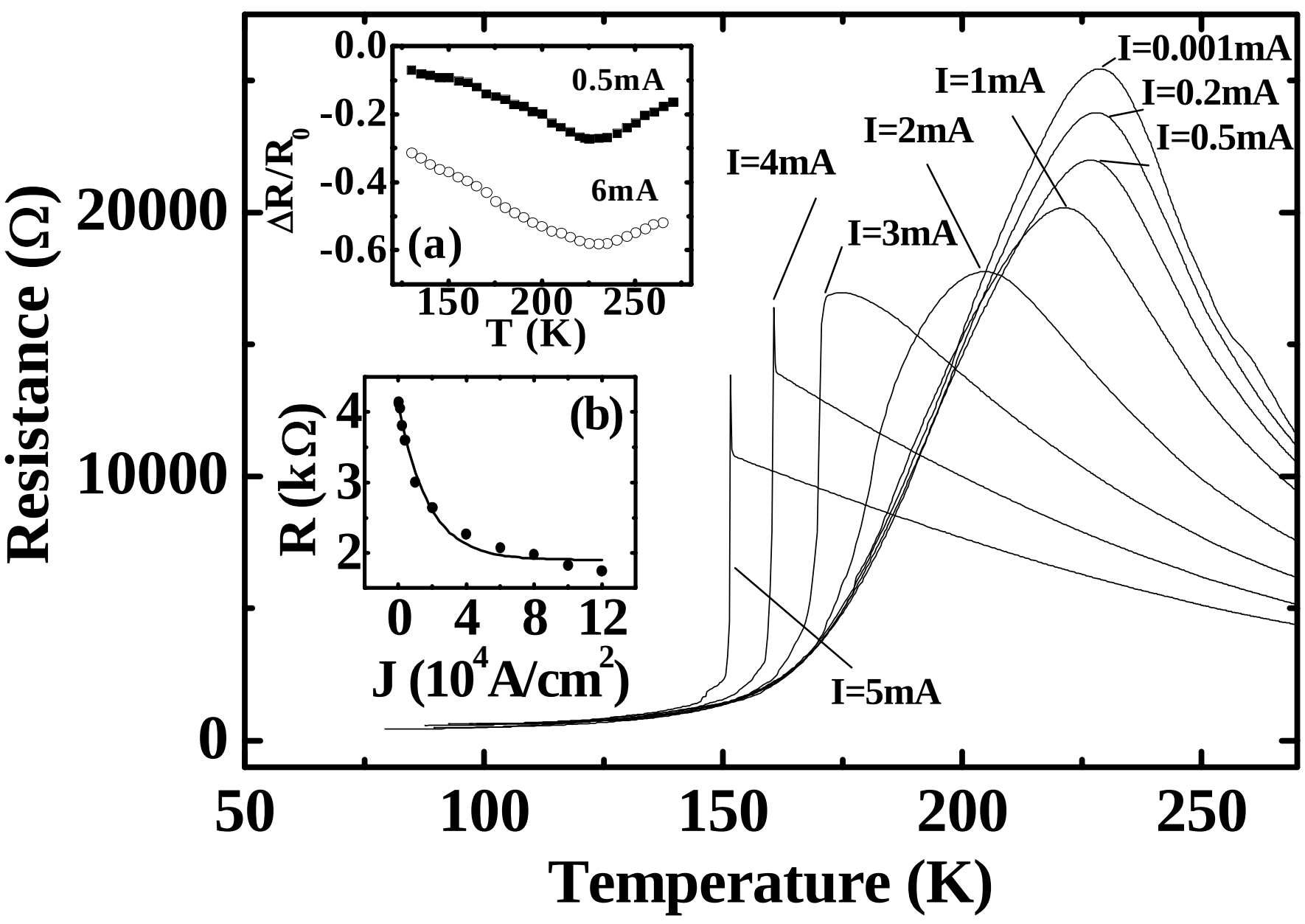

Fig.2

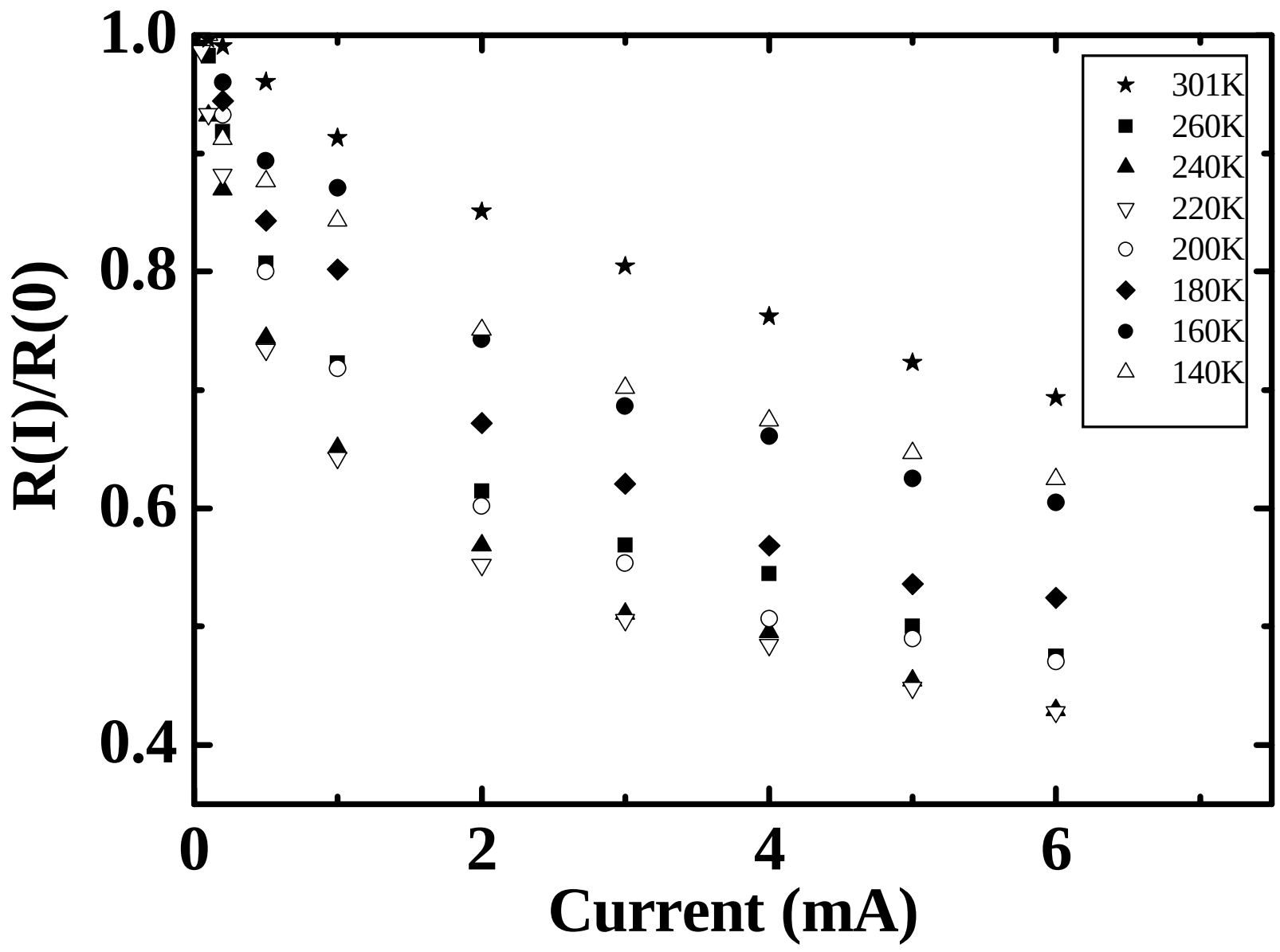

Fig.3

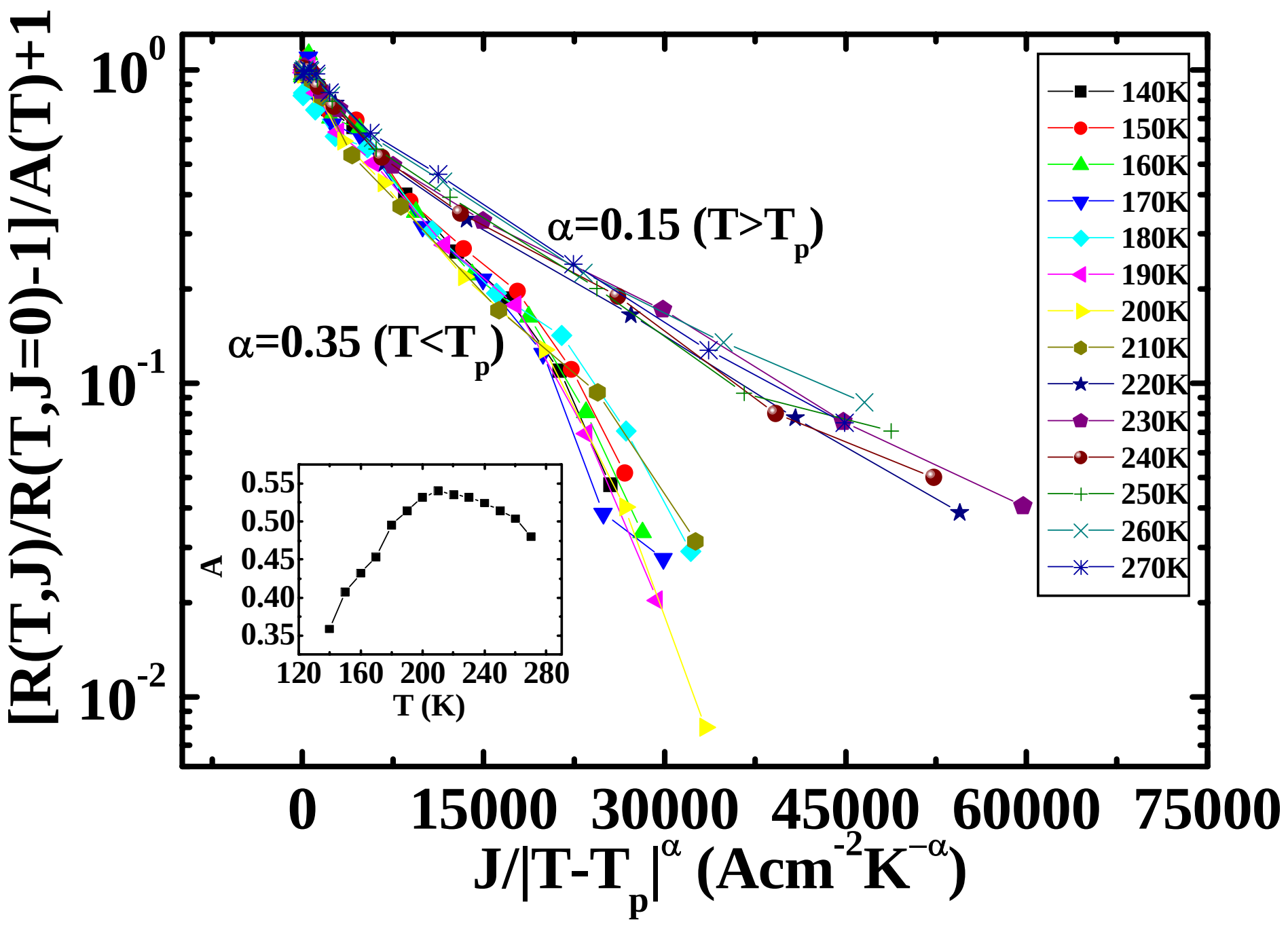

Fig.4